\documentstyle[12pt]{article}
\begin{document}
%
%
\title{\bf 
Quantum Phenomenology\\   
with the Path Integral Approach}
\author{
\normalsize
T. Calarco,$^{1,2}$ R. Onofrio,$^{1,3}$ C. Presilla,$^4$ 
and L. Viola$^{1,3}$\\
\normalsize
$^1$Dipartimento di Fisica ``G. Galilei'', Universit\`a di Padova,\\
\normalsize
Via Marzolo 8, Padova, Italy 35131\\
\normalsize
$^2$INFN, Laboratori Nazionali di Legnaro, Legnaro, Italy 35020\\
\normalsize
$^3$INFN, Sezione di Padova, Padova, Italy 35131\\
\normalsize
$^4$Dipartimento di Fisica and INFN, Universit\`a di Roma ``La Sapienza'',\\
\normalsize
Piazzale A. Moro 2, Roma, Italy 00185}
\date{ }
\maketitle
%
%
\begin{abstract}
A quantum measurement model based upon restricted path-integrals allows 
us to study measurements of generalized position in various one-dimensional 
systems of phenomenological interest. 
After a general overview of the method we 
discuss the cases of a harmonic oscillator,
a bistable potential and two coupled systems, briefly illustrating their 
applications.
\end{abstract}
\leftline{PACS: 03.65.Bz, 06.30.-k, 74.50.+r}
\leftline{Keywords: Quantum measurements, Bell inequalities}
\vspace{0.5cm}
\date{ }
\maketitle
%
%
Recent efforts in high precision experiments have shown that the fundamental 
limitations to a measurement due to quantum mechanics play a crucial
role to develop more and more sensitive instruments \cite{BRAG}. 
As a by-product quantum measurement theory, a topic
quite isolated from the frontier of physics some decades ago, 
has been revived and new experiments have been 
proposed for its better understanding.
Models of quantum measurement theory comparable to 
the outcomes of the experiments are therefore welcome.
Here we  describe some
of the results obtained using one of these models, originally developed in
\cite{ME1} (see also \cite{ME2} for a complete account of the approach). 
The goal of this paper is twofold: to discuss specific examples 
of systems in which quantum measurement theory plays a crucial 
role and to show that they are all described within the {\it unified} 
context of the model we use.
For reasons of space we do not deal here with 
a comparison of our approach to the other ones 
described elsewhere in these Proceedings \cite{GEN}.
Moreover, we remand the interested reader to the mentioned references 
of our papers for more details. 

\section*{A model for impulsive measurements}

In classical mechanics the effect of a sequence of measurements 
on the subsequent dynamics is negligible.
In quantum mechanics instead the execution of a measurement, 
either continuous or stroboscopic, influences the system. 
In this last case it is therefore important to include the 
effect of an actual measurement on the dynamics of the observed system. 
This has been obtained by Mensky 
\cite{ME1,ME2} through the path-integral approach (other similar 
approaches have been proposed in \cite{GEN}). 
In this framework the propagator of a system 
described by a Lagrangian ${\cal L}(x(t),\dot{x}(t),t)$, and undergoing
a continuous measurement of its position between the 
times 0 and $\tau$, with result $a(t)$ and instrumental 
uncertainty $\Delta a$, is written as a weighted path-integral:
\begin{equation}
\label{MENS}
K_{[a]}(x^{\prime\prime},\tau;x^{\prime},0)=
\int_{x(0)\equiv x^{\prime}}^{x(\tau)\equiv x^{\prime\prime}}
 {\cal D}[x(t)]\exp \left\{\frac{i}{\hbar}\int_0^{\tau}{\cal L}(x(t),
\dot{x}(t),t)dt \right\}w_{[a]}[x],
\end{equation}
where
\begin{equation}
\label{PESOW}
w_{[a]}[x]=\exp \left\{-\frac{1}{2\Delta a^2 \tau}
\int_0^{\tau} [x(t)-a(t)]^2 dt \right\}.
\end{equation}
The most natural way to represent an impulsive measurement 
of position at time 0 is as limit of a continuous one for infinitesimal 
time intervals.  In this approximation,
\begin{equation}
\label{IMP}
\lim_{\tau\to 0}
K_{[a(t)]}(x^{\prime\prime},\tau;x^{\prime},0)=
K_{a}(x^{\prime\prime},x^{\prime}) = e^{-\frac{(x^{\prime}-a)^2}{2\Delta a^2}}
K(x^{\prime\prime},0;x^{\prime},0)
\end{equation}
where $a=a(0)$.
Let us suppose that
$\psi(x,t)$ be the wavefunction of a system subjected to an impulsive 
measurement of position at time $t$, with result $a$.
Since  $K(x^{\prime\prime},0;x^{\prime},0)=
\delta(x^{\prime\prime}-x^{\prime})$,
 it follows
\begin{equation}
\label{PSII}
\psi(x,t^+)  =  R_{[\psi]}(a)
\int_{-\infty}^{+\infty}dyK_{a}(x,y)\psi(y,t^-)=
R_{[\psi]}(a)\exp\left\{-\frac{(\hat{x}-a)^2}
{2\Delta a^2}\right\}\psi(x,t^-).
\end{equation}
where $R_{[\psi]}(a)$ is a renormalization constant. 
The effect of the measurement is therefore a Gaussian 
filtering around the measurement result.
This is similar to the usual measurement theory of von Neumann. 
The latter is simply recovered by choosing for the measurement 
operator the discontinuous and therefore less realistic form 
\begin{equation}
\label{misvn}
\hat{w}^{v.N.}_a\propto 
\theta(\hat{x}-[a-\Delta a])\theta([a+\Delta a]-\hat{x}).
\end{equation}
For a sequence of impulsive 
measurements the Gaussian reduction (\ref{PSII}) 
is alternatively followed by free evolution periods $\Delta T$ 
(quiescent time).
The probability that the $N^{th}$ measurement gives a result $a_N$, 
when the results of the previous $N-1$ ones are known, is expressed 
through
\begin{equation}
\label{PROBEPSI}
P_{a_1,a_2,..,a_{N-1}}(a_N)\equiv\frac{\left|\left\langle\psi_{a_1,a_2,..,a_N}
(t_N^+)\right|\left.
\psi_{a_1,a_2,..,a_N}(t_N^+)\right\rangle\right|^2}{\displaystyle{
\int_{-\infty}^{+\infty}
\left|\left\langle\psi_{a_1,a_2,..,a_N}(t_N^+)\right|\left.
\psi_{a_1,a_2,..,a_N}(t_N^+)\right\rangle\right|^2d a_N}}.
\end{equation}
which allows one to evaluate the effective uncertainty defined as
\begin{equation}
\Delta a_{\it eff}^2=
{2 \int (a_N-\tilde{a})^2 ~ 
P_{a_1,a_2,..,a_{N-1}}(a_N)~da_N\over
 \int P_{a_1,a_2,..,a_{N-1}}(a_N)~da_N}. 
\label{DAEFFTH}
\end{equation}
where $\tilde{a}$ is the most probable measurement result.
The effective uncertainty $\Delta a_{eff}$ 
expresses  the spreading of the possible
measurement results \cite{MEOP2} and is equal to $\Delta a$, the 
instrumental error, in the classical limit. 
In a quantum regime $\Delta a_{\it eff}$ 
is always larger or equal to $\Delta a$.
A convenient situation for evaluating the $\Delta a_{eff}$ is 
available in the case of impulsive measurements 
when energy eigenstates and eigenvalues of the system are known. 
In such a case the evolution of the state during the 
quiescent time is obtained once an expansion in terms of energy 
eigenstates of the output of a measurement is made and  
the numerator of (\ref{PROBEPSI}) can be rewritten as 
\begin{equation}
\label{PSIO}
\left|\left\langle\psi_{a_1,a_2,..,a_N}(t_N^+)\right|\left.
\psi_{a_1,a_2,..,a_N}(t_N^+)\right\rangle\right|^2 = 
\left(\sum_{m=1}^\infty\left|c_m^{(N)}\right|^2\right)^2
\end{equation}
where $ c_m^{(N)}$ is the projection coefficient on the $m^{th}$ eigenstate 
$|m\rangle$ at the $N^{th}$ measurement.
Thus from the energy eigenstates expansion of the initial wavefunction 
it is possible to derive the effective uncertainty 
\begin{equation}
\label{DPHIEFF}
\Delta a_{\it eff}^2=
\frac{\displaystyle{2 \int_{-\infty}^{+\infty}(a_N-\tilde{a})^2
\left(\sum_{m=1}^\infty\left|\sum_{l=1}^\infty 
B_{ml}^N(a_0,\ldots,a_{N-1},a_N)
c_l^{(0)}\right|^2\right)^2d a_N}}
{\displaystyle{\int_{-\infty}^{+\infty}
\left(\sum_{m=1}^\infty\left|\sum_{l=1}^\infty 
B_{ml}^N (a_0,\ldots,a_{N-1},a_N)
c_l^{(0)}\right|^2\right)^2d a_N}},
\end{equation}
where 
\begin{equation}
\label{CNM}
c_m^{(N)}\stackrel{\rm def}{=}\sum_{l=1}^\infty B_{ml}^Nc_l^{(0)},
\end{equation}
\begin{equation}
\label{BML}
B_{ml}^N(\Delta T,\Delta a,a_1,a_2,..,a_N)\stackrel{\rm def}{=}\langle m|
\hat{w}_{a_N}\left(\prod_{j=1}^N
e^{-\frac{i}{\hbar}\hat{H}\Delta T}\hat{w}_{a_{N-j}}\right)|l\rangle.
\end{equation}
For quantum measurements on harmonic oscillators, having energy levels equally 
spaced, the optimal quiescent time is half the  oscillation period and all 
can be evaluated without approximations.
This is already known for the limit case of instantaneous  and 
infinite accuracy measurements 
as seen using the canonical approach: the commutator for the position 
at different times gives \cite{CAVES} 
\begin{equation}
\label{COMMU}
 [\hat{x}(t+\Delta T), \hat{x}(t)]={i\hbar \over {m\omega}} \sin{\omega 
\Delta T}
\end{equation}
which implies that for two instantaneous measurements of position 
spaced by multiples of half period of oscillation the observable is 
quantum nondemolition. The path integral formalism allows us 
to extend such a result to finite accuracy and finite duration
measurements, as shown in \cite{MEOP2}. It turns out that the
effective uncertainty holds values very close to the instrumental 
uncertainty only for quiescent times which are multiples of half period of
the harmonic oscillator. The optimal measurements of position for 
a harmonic oscillator are also known in literature as quantum 
nondemolition measurements \cite{BRAG,CAVES}, and are quite important 
for the detection of small displacements 
like pulses of gravitational waves of astrophysical origin.

\section*{Quantum measurements in SQuIDs}

Superconducting Quantum Interferometer Devices (SQuIDs) have been 
proposed to test macrorealism versus quantum mechanics. 
A global understanding of quantum mechanics and its relationship to the 
classical limit requires to extend its validity to the macroscopic world. 
In this domain the conflict between its structure and the classical 
sense of physical reality cannot be overcome \cite{BELL} and experiments 
aimed to compare the predictions of the two worldviews are crucial. 
By realism here we mean that ``a macroscopic system with two or more 
macroscopically distinct states available to it will at all times {\it be} 
in one or the other of these states'' ({\it macroscopic realism}) as well as 
that ``it is possible, in principle, to determine the state of the system with 
arbitrarily small perturbation on its subsequent dynamics'' ({\it non-invasive 
measurability at the macroscopic level}) \cite{LEGG}.
This leads to a quantitative test by introducing 
inequalities between correlation functions of observables of a macroscopic 
system, a superconducting quantum interferometer device 
subjected to a sequence of repeated measurements of magnetic flux. 
The proposal of Leggett and Garg \cite{LEGG} has been 
criticized due to the limitations given by quantum 
mechanics to the accuracy obtainable in a set of 
repeated measurement of the same observable \cite{BAL}. 
However, measurement schemes following that path have been proposed 
both using a set of SQuIDs and two-level mesoscopic systems \cite{TE}.
We have studied quantitatively optimal strategies for repeated measurements of 
magnetic flux in bistable potentials which can schematize the SQuID 
behaviour \cite{CALARCO}. 
It turns out that measurement strategies must be chosen in a particular way to 
minimize the influence of the previous measurements on 
the state of the observed system: quantum nondemolition strategies 
for the measurement of the magnetic flux in a SQuID have to be implemented.
The magnetic flux in a SQuID is schematized, if the coupling 
to the external environment can be neglected, through the effective 
potential
\begin{equation}
V(\varphi)=-{\mu\over 2} \varphi^2+{\lambda \over 4} \varphi^4
\label{SOMBRER}
\end{equation} 
where $\varphi$ is the trapped magnetic flux, a generalized coordinate 
describing the system,
$\mu$ and $\lambda$ are parameters associated to the superconducting circuit.
This allows us to describe the system in terms of pure states $\psi(\varphi)$.
Since general arguments exist on the fundamental noise introduced by 
any linear amplifier \cite{HEFF}, the problem of the measurement 
of flux in a superconducting circuit is independent upon the 
detailed scheme used to detect the quantum state of the SQuID.
 As already discussed in \cite{MEOP2} the optimality of the measurement 
is dictated by the spectral properties of the system.
The asymptotic collapsed wavefunction can be expanded in terms 
of the eigenstates of the system and an optimal measurement 
is obtained provided that the quiescent time is commensurable 
to the characteristic times for the wavefunction reformation
\begin{equation}
\label{tij}
T_{ij}\equiv\frac{h}{|E_i-E_j|},
\end{equation}
where $E_i$, $E_j$ are the energy eigenvalues which have maximal 
projections on the asymptotic state.
The two eigenstates which maximally contribute to the wavefunction 
reformation after tunneling are those corresponding to 
the first two eigenvalues, whose splitting is related to
the tunneling period. 
The effective error $\Delta \Phi_{\it eff}$ reached after a series
of mesurement pulses has minima
close to $\Delta \Phi$ when the quiescient time $\Delta T$ is a multiple 
of the tunnelling period $T_{12}$.
Unless the measurements are repeated with this periodicity noise due to the 
measurement process is fed into the system 
affecting the following measurements. 
In the case of the potential of Eq. (\ref{SOMBRER})
we have numerically evaluated the eigenstates using a selective 
relaxation algorithm \cite{PRETA}.
The predictability of the measurement outcome 
is affected by the uncertainty due to the previous measurements.
One can minimize this by simply using optimal quiescent times which are 
analytically determined through (\ref{tij}) 
once the eigenvalues and the eigenstates are given. \\
The general hypotheses under which we have obtained this result 
allows us to conclude that a direct experimental test of temporal 
Bell-type inequalities in principle  
should remain possible even in a pure quantum mechanical framework, without 
any classical assumption on the measurement process such as noninvasivity 
\cite{LEGG}. 
However, this happens only if the repetition times
between consecutive measurements which violates Bell 
inequalities are compatible with the optimal quiescent times.  
Otherwise either the inequalities are not violated or 
quantum noise is introduced in the successive measurement, making it invasive
\cite{CALARCO}.
A preliminary analysis on which we will refer in the future 
shows that this is the case. 
The region of violation occurs when the two quiescent times required to 
have a set of three possible measurements are both 
of the order of half of the tunnelling period. 
Unfortunately for such periodicities the 
effect of the measurement is such that the spreading on the 
possible results induced by the measurement does not allow one  
to distinguish between the two wells of the bistable potential.
As a consequence violation of temporal Bell inequalities cannot be observed.

\section*{Quantum measurements in Penning traps}

Penning traps consist of a combination of static magnetic and electric 
fields where a single charged particle, for instance an electron, can be 
stored for long times \cite{BROWN}. 
A magnetic field confines the motion in a cylinder whose size 
is determined by the strenght of the magnetic field itself. 
A further confinement along the axis of the cylinder is obtained 
through a quadrupole electrostatic field creating a harmonic 
force parallel to the magnetic field.
Thus the motion of the electron is a combination of a harmonic motion 
along the axis of the cylinder and a circular motion, called cyclotron 
motion, completely decoupled in the non-relativistic limit. 
They become coupled, with the spin too, by including relativistic 
corrections. 
In this case the Hamiltonian of an electron in a Penning trap is
\begin{eqnarray}
H_{sys}=\hbar \omega_c a_c^+ a_c 
\left\{ 1-1/2 {\left( {\omega_z \over \omega_c} \right)}^2 
- {\hbar \omega_c \over {2m_0c^2}} \right\} 
- {(\hbar \omega_c)^2 \over {2m_0c^2}}{(a_c^+ a_c)}^2 + 
\nonumber\\
+{p_z^2 \over {2m_0}} + {m_0 \omega_z^2 z^2\over 2} 
- {1 \over {2m_0 c^2}} {\left( {p_z^2 \over 2 m_0} \right)}^2 
+ {\hbar \omega_z \sigma_z\over 4} \left[g-{\hbar \omega_c \over {m_0 c^2}}
\left(1+{g-1\over 2}
{\left({\omega_z \over \omega_c}\right)}^2\right)\right]-
\nonumber\\
- {(\hbar \omega_c)^2\over {2m_0 c^2}} \left[1+{g-1\over 2}
{\left(\omega_z \over \omega_c \right)}^2 \right] \sigma_z a_c^+ a_c 
- {p_z^2 \over {2m_0^2c^2}} \hbar \omega_c a_c^+ a_c 
-g {p_z^2 \over {8m_0^2c^2}}\hbar \omega_c \sigma_z 
\end{eqnarray}
\noindent
where $a_c$ and $a_c^+$ are the annihilation and 
creation operators associated to the cyclotron motion, $z$, $p_z$ are 
position and momentum for the axial motion and $\sigma_z$ is the 
projection of the spin on the $z$-axis.
The interactions among the three degrees of freedom 
have small coupling constants proportional to $c^{-2}$. 

Recently it has been proposed to measure the energy 
of the cyclotron motion of an electron in a Penning trap 
in a quantum nondemolition way \cite{TOMBESI}. 
A measurement of the energy associated to the cyclotron motion is obtained 
through the measurement of the axial motion of the trapped particle
provided that the spin is preassigned. 
A detailed analysis of the dynamics of such a measurement is still 
missing, and the model discussed here can be applied to this situation.
We note two peculiarities of such an application. Firstly, one is 
dealing with the case of non-quadratic terms in the Hamiltonian and 
numerical analysis is mandatory. Secondly, this is an example 
of an {\it indirect measurement},  
in which the informations obtained on one degree of freedom allow one 
to measure some other quantities related to another 
degree of freedom coupled to the first
(in this particular case the cyclotron energy through 
the direct measurement of spin and axial motion of the electron). 
The simplest prototype of systems in which indirect measurements 
are defined could consist of two coupled harmonic oscillators with only 
one subjected to measurement. 
In the particular case of position measurements 
the corrisponding Hamiltonian could be written as
\begin{equation}
\label{DOUBLE}
H_{eff}={p_1^2 \over {2m_1}}+{m_1 \omega_1^2 x_1^2 \over 2}+\gamma x_1 x_2+
{p_2^2 \over {2m_2}}+{m_2 \omega_2^2 x_2^2 \over 2}+ 
{i \hbar\over {\Delta a_1^2 \tau}}[x_1(t)-a_1(t)]^2 
\end{equation}
where the two oscillators are linearly coupled with strenght $\gamma$, 
and the measurement term directly affects only the first oscillator. 
It is easy to realize that, due to the coupling, a restriction of 
the paths in the coordinate space of the first oscillator will induce 
a restriction of the paths also in the second oscillator, therefore defining 
an indirect effective uncertainty $\Delta a_{2\it eff}$.
A perturbative evaluation of such a quantity in the case of 
Eq. (\ref{DOUBLE})
is ongoing and we will refer on it in the future.\\
The interest of the above 
problem goes beyond the Penning traps application, and involves
topics such as gravitational wave antennae coupled to a transducer 
\cite{RAPA1} 
and quantum measurements of energy in microwave cavities \cite{BRUNE}. 

\vspace{0.5cm}
We acknowledge M. B. Mensky for collaboration and 
stimulating discussions. 
%
%

\end{document}